\documentstyle[aps,epsf,psfig]{revtex}
\newcommand{\be}{\begin{equation}}
\newcommand{\ee}{\end{equation}}
\newcommand{\ba}{\begin{eqnarray}}
\newcommand{\ea}{\end{eqnarray}}
\newcommand{\Tr}{\mathrm{Tr}}
\def\reff#1{(\ref{#1})}
\newcommand{\1}{1\!\!\!\bot}
\def\gtapprox{\raisebox{0.45ex}{$\,\,>$}\raisebox{-0.7ex}{$\!\!\!\!\!\sim\,\,$}}
\def\ltapprox{\raisebox{0.45ex}{$\,\,<$}\raisebox{-0.7ex}{$\!\!\!\!\!\sim\,\,$}}
\begin{document}
%
%
\title{ { \normalsize \hfill \parbox{28mm}{BI-TP 99/03} }\\[40mm]
          \vspace{-2cm} Infrared behavior of the gluon propagator
                        in lattice Landau gauge: \\
                        the three-dimensional case }

\author{Attilio Cucchieri\thanks{E-mail address: 
       {\tt attilio@physik.uni-bielefeld.de}.}}
\address{{\small Fakult\"at f\"ur Physik, Universit\"at Bielefeld, 
                 D-33615 Bielefeld, GERMANY}}
\date{\today}
\maketitle
\begin{abstract}
We evaluate numerically the three-momentum-space gluon propagator in the lattice
Landau gauge, for three-dimensional pure-$SU(2)$ lattice gauge theory with
periodic boundary conditions. Simulations are done for nine different values
of the coupling $\beta$, from $\beta = 0$ (strong coupling) to
$\beta = 6.0$ (in the scaling region), and for lattice sizes up to $V = 64^3$.
In the limit of large lattice volume we observe, in all cases, a
gluon propagator decreasing for momenta smaller than a constant value
$p_{dec}$. From our data we estimate
$p_{dec} \approx 350 \, \mbox{MeV} $.
The result of a gluon propagator decreasing in the infrared limit
has a straightforward interpretation as resulting from the proximity of the
so-called first Gribov horizon in the infrared directions.
\end{abstract}
%
%
\section{Introduction}

The infrared behavior of the gluon propagator in lattice Landau gauge has been
the subject of several numerical studies
\cite{MO2,Gup,Nakamura,Aiso,gut,par,Athesis,Attilio1,Attilio2,Naka}.
In fact, although
this propagator is a non-gauge-invariant quantity, the study of its
infrared behavior provides a powerful tool for increasing our understanding of
QCD, and for gaining insight into the physics of confinement in non-abelian
gauge theories (see for example
\cite{Rober}).
In particular, the infrared behavior of the
gluon propagator can be directly related
\cite{West}
to the behavior of the Wilson loop at large separations, and to the
existence of an area law.
 
On the lattice, the Landau gauge condition is imposed
\cite{MO2,W2}
by finding a gauge transformation which
brings the functional ${\cal E}_{U}[ g ]$,
defined in eq.\ \reff{eq:Etomin} below, to a minimum. A lattice configuration
satisfying this minimizing condition belongs to
the region $\Omega$ of transverse
configurations, for which the Faddeev-Popov operator is nonnegative
\cite{DZ,Z4,Z1}.
This region is delimited by the so-called first Gribov horizon, defined as
the set of configurations for which the smallest non-trivial eigenvalue
of the Faddeev-Popov operator is zero. (The Faddeev-Popov operator has a
trivial null eigenvalue, corresponding to a constant eigenvector.) 

The restriction of the configuration space
to the region $\Omega$ implies a {\em rigorous} inequality
\cite{DZ,Z4,Z1}
for the Fourier components of the gluon field $ A_{\mu}(x) $.
From this inequality, which is a
consequence of the positiveness of the Faddeev-Popov operator only, it
follows that the region $\Omega$ is bounded by a certain
ellipsoid $\Theta$. This bound implies proximity of the first Gribov horizon
in infrared directions, and consequent suppression of the low-momentum
components of the gauge field, a result already noted by Gribov in Reference
\cite{Gr}.
This bound also causes a strong suppression of the gluon propagator
in the infrared limit (i.e.\ for momentum $p \to 0$). In fact, Zwanziger proved
\cite{Z1,Z2}
that, in four dimensions and in the infinite-volume limit, the gluon propagator
is less singular than $p^{-2}$ in the infrared limit and that, very likely, it
{\em does} vanish in this limit. More precisely, in Reference
\cite{Z2}
it was proven that, in the infinite-volume limit,
the ``gluon propagator'' $D(H,\,p)$ goes to zero as $p \to 0$
for {\em almost every} $H$. Here $H$ should be interpreted as the strength of a
spatially-modulated magnetic field coupled to the gluon field $ A_{\mu}(x) $,
and the standard gluon propagator is obtained for $H = 0$.
A similar result holds in three dimensions: one obtains
\cite{Z1,Z2}
that, in the infinite-volume limit,
the gluon propagator must be less singular than $p^{-1}$ as $p \to 0$ and that,
very likely, it vanishes in the infrared limit.

A gluon propagator vanishing in the infrared limit was also
found --- in four dimensions --- by Gribov
\cite{Gr}.
More precisely, he obtained the expression $ p^{2} / ( p^{4} \,+\, \gamma ) $,
where the mass scale $\gamma^{1/4}$ arises
when the configuration space is restricted to the region $\Omega$.
This propagator agrees with the zeroth-order perturbative prediction
$p^{- 2}$ at large momenta, but gives a null propagator at $p = 0$. The
mass scale $\gamma^{1/4}$ marks the transition point between the perturbative
and the nonperturbative regimes. A propagator
that is a generalization of the one obtained by Gribov has also been introduced
in Reference
\cite{Stingl}
as an Ansatz for a non-perturbative solution of the gluon Dyson-Schwinger
equation (DSE).

Let us notice that a gluon propagator vanishing in the infrared limit is
in complete contradiction with the $p^{- 4}$ singularity obtained when the gluon
DSE is {\em approximately} solved in the infrared limit
\cite{Rober,BPenn}.
However, a recent study
\cite{smekal}
has shown that this singularity is obtained only if
the ghost contributions to the gluon DSE are neglected. In fact, when these
contributions are included, the gluon propagator vanishes in the infrared limit
\cite{smekal},
in qualitative agreement with References
\cite{Z1,Gr,Z2}.

 
\vspace{0.3cm}

In this paper we present the {\em first} numerical study of the
infrared behavior of the gluon propagator in three
dimensions.\footnote{Preliminary results have been reported in
\protect\cite{lat9798}.}
As explained in References
\cite{3dprimo,vari3d,Teper},
nonabelian gauge theories in three dimensions are similar to
their four-dimensional counterparts, and results obtained in the
three-dimensional case can teach us something about the more
realistic four-dimensional theories.
Of course, the advantage of using a three-dimensional lattice is the possibility
of simulating lattice sizes larger than those used in the four-dimensional case.
This is particularly important in the study of the gluon propagator since
Zwanziger's prediction
of an infrared-suppressed gluon propagator is valid only
in the infinite-volume limit.

We recall that in some recent numerical studies (in four dimensions)
\cite{gut,par}
a sensible change in the infrared behavior of the gluon propagator
has been observed for momenta smaller than a turn-over value $p_{to}$,
in agreement with the prediction of a gluon propagator less
singular than $p^{-2}$ in the infrared limit. Also, the numerical data
obtained in Reference
\cite{Nakamura}
have been successfully fitted by a Gribov-like formula
\cite{Aiso}.
Finally, in References
\cite{Athesis,Attilio1,Attilio2}
we have observed --- for the $SU(2)$ group, in four dimensions,
in the strong-coupling regime and in the limit of large lattice
volume --- a gluon propagator {\em decreasing} as the momentum
goes to zero. A similar result has also been obtained recently 
\cite{Naka}
for the $SU(3)$ group in the strong-coupling regime, both in the quenched and in
the unquenched case. Let us notice that Zwanziger's predictions
\cite{Z1,Z2}
for the gluon propagator are {\em $\beta$-independent}: in fact,
they are derived only from the positiveness of the Faddeev-Popov
operator when the lattice Landau gauge is imposed. Thus, results in
the strong-coupling regime (i.e.\ $\beta \to 0$) are a valid test
of these predictions. Nevertheless, it is important to extend
this result to higher values of $\beta$, possibly up to the scaling
region. Of course, as $\beta$ increases, one needs to consider
larger lattice sizes in order to probe the infrared behavior of
the gluon propagator. Moreover, as said above, the prediction we want to
test applies only in the infinite-volume limit. In four dimensions
it was found
\cite{Athesis,Attilio1,Attilio2}
that the lattice size at which an infrared-decreasing gluon propagator
starts to be observed increases with the coupling.
This makes
practically unfeasible, with present computational resources, to study
numerically the infrared behavior of the gluon propagator in the
four-dimensional case and at values of $\beta $ in the scaling
region. Our hope is that in this work, by studying the three-dimensional
case, we can consider lattice volumes that are large enough to allow a
decreasing gluon propagator to be observed not only in the strong-coupling
regime but also in the scaling region.


\section{Definitions and Notations}
\label{sec:propag}
 
We consider a standard Wilson action for $SU(2)$ lattice gauge theory
in $3$ dimensions
\be
S[ U ] \equiv
   \frac{4}{a \, g^{2}} \, \frac{1}{2} \sum_{\mu, \nu = 1}^{3}\,
      \sum_{x}\, \biggl\{ \, 1 \, - \frac{\Tr}{2} \left[\
U_{\mu}(x) \, U_{\nu}(x + e_{\mu}) \,
U_{\mu}^{-1}(x + e_{\nu}) \, U_{\nu}^{-1}(x) \right]
\, \biggr\}
\;\mbox{,}
\label{eq:Sdef}
\ee
where $U_{\mu}(x) \in SU(2)$ are link variables, $g$ is the bare
coupling constant, $a$ is the lattice spacing, and $e_{\mu}$ is a unit vector
in the positive $\mu$ direction. We assume periodic boundary conditions.
For the matrices $U \in SU(2)$ we use the parametrization
$ U\,\equiv \, U_{0}\,1\!\!\!\bot\,+\,i\,{\vec{\mbox{$U$}}}\cdot
{\vec{\mbox{$\sigma$}}} $,
where $\1$ is the $2 \times 2$ identity matrix, the components of
${\vec{\mbox{$\sigma$}}} \equiv ( \sigma^{1} \mbox{,} \, \sigma^{2} \mbox{,} \,
\sigma^{3} ) $ are the Pauli matrices, $U_{0} \in \Re$, ${\vec{\mbox{$U$}}} \in
\Re^{3}$ and $ \,U_{0}^{2} + {\vec{\mbox{$U$}}} \cdot {\vec{\mbox{$U$}}} = 1$.
Notice that in eq.\ \reff{eq:Sdef} the lattice spacing $ a $ is necessary in
order to make the action $ S[U] $ dimensionless;
in fact, in the three-dimensional case
\cite{3dprimo,vari3d,Teper},
the coupling $g^{2}$ has dimension of mass, and in order to obtain a
dimensionless lattice coupling we have to set $\beta \equiv 4 / (a\,g^{2}) $.

We define the gauge field $ A_{\mu}(x) $,
which belongs to the ${\mathcal SU}(2)$ Lie algebra, as
\be
A_{\mu}(x) \,\equiv\, \frac{1}{2}\,
    \left[\,U_{\mu}(x)\,-\,U_{\mu}^{\dagger}(x)\,\right]\,
=\,i\,{\vec{\mbox{$U$}}}_{\mu}(x)\cdot
     {\vec{\mbox{$\sigma$}}}
\;\mbox{.}
\label{eq:Amux}
\ee
We also define
\be
A_{\mu}^{b}(x) \,\equiv\, \frac{\Tr}{2 i}
   \left[ \, A_{\mu}(x) \, \sigma^{b} \, \right]\,
=\,U_{\mu}^{b}(x)
\;\mbox{,}
\label{eq:Amuxb}
\ee
where $\sigma^{b}$ is a Pauli matrix. Note that $ a^{-1} A_{\mu}^{b}(x) $
approaches
$( 1 / 2 ) \, g \, [ A^{(cont)} ]^{b}_{\mu}(x) \,$ in the continuum limit, where
$[ A^{(cont)} ]^{b}_{\mu}(x) $ is the
(unrenormalized) conventional vector potential.
 
 
\vspace{0.3cm}

In order to fix the lattice Landau gauge we look for a local
minimum\footnote{Here we do not consider the problem
of searching for the {\em absolute} minimum of the functional
$\,{\cal E}_{U}[ g ]\,$, which defines the so-called {\em minimal Landau gauge}
\cite{Z2}.
In fact, as stressed in the Introduction, the prediction of
an infrared-suppressed gluon propagator is valid for any configuration in the
region $\Omega$, i.e.\ for local as well absolute minima of the functional
$\,{\cal E}_{U}[ g ]\,$.
\protect\label{foo:Gribov}}
of the functional
\cite{MO2,W2}
\be
{\cal E}_{U}[ g ]\,\equiv\,1\,-
       \frac{1}{3\,V}\,\sum_{\mu = 1}^{3}\,
  \sum_{x}\,\frac{\Tr}{2}\,
\left[\,g(x)\,U_{\mu}(x)\,g^{\dagger}(x + e_{\mu})\,\right]
\;\mbox{,}
\label{eq:Etomin}
\ee
where $\,g(x)\in SU(2)\,$ are site variables, and $\, V \equiv N_{s}^{2}\,
N_{t}\,$ is the lattice volume. (Here $ N_{s}$ is the number of lattice
sites in the two spatial directions, and $ N_{t}$ is the number of lattice
sites in the time direction.) If the configuration
$\{ U_{\mu}\left(x\right) \}$ is a stationary point of the functional
${\cal E}_{U}[ g ]$ then
\cite{W2}
the lattice divergence of $ A_{\mu}^{b}(x) $ is null, namely
\be
\left(\nabla\cdot A \right)^{b}(x) \equiv
  \sum_{\mu = 1}^{3} \, \left[ A_{\mu}^{b}(x) -
                  A_{\mu}^{b}(x - e_{\mu}) \right]
               \, = \, 0 
\qquad \qquad \;\;\; \forall \; \; \; x \mbox{,} \; \; b \;\mbox{.}
\label{eq:diverg0}
\ee
This is the lattice formulation of the usual (continuum) Landau gauge-fixing
condition. By summing eq.\ \reff{eq:diverg0} over the components
$x_{\mu}$ of $x$ with $\mu \neq \nu$, for fixed $\nu$,
and using the periodicity of
the lattice, it is easy to check
\cite{MO2}
that if the Landau gauge-fixing condition
is satisfied then the quantities
\be
Q_{\nu}^{b}(x_{\nu}) \, \equiv \, \sum_{\mu \neq \nu} \,
     \sum_{x_{\mu}} \, A_{\nu}^{b}(x)
\label{eq:charges}
\ee
are constant, i.e.\ independent of $x_{\nu}$.


\section{Gluon propagator on the lattice}

The lattice space-time gluon propagator is given by
\be
D_{\mu\, \nu}^{b\, c}( x - y )
\equiv \langle\,A^{b}_{\mu}(x)\, A^{c}_{\nu}(y)\,\rangle
\;\mbox{.}
\ee
To go to momentum space we can use Formula (3.1a) in Reference
\cite{Z1}
and obtain
\ba
D(0)& \equiv& \frac{1}{9 V} \sum_{\mu\mbox{,}\,b}\,\langle\,
  \left[\,\sum_{x}\,A_{\mu}^{b}(x)\,\right]^{2} \rangle
\label{eq:D0def} \\
D(k) & \equiv & \frac{1}{6 V} \sum_{\mu\mbox{,}\,b}\,\langle\,
\left\{\,\left[\,\sum_{x}\,A_{\mu}^{b}(x)\,
\cos{( 2 \pi k \cdot x )}\,\right]^{2} 
 + \left[\,\sum_{x}\,A_{\mu}^{b}(x)\,
\sin{( 2 \pi k \cdot x )}\,\right]^{2}
\, \right\} \,\rangle
\;\mbox{.}\;\;\;\;\;\;
\label{eq:Dkdef}
\ea
Here $\mu$ goes from $1$ to $3$, and $k$ has components $(k_{x}\mbox{,}\,
k_{y}\mbox{,}\, k_{t})$. In our simulations we consider the values
$k_{x}\,N_{s} = k_{y}\,N_{s} = 0\mbox{,}\,1$ and $k_{t}\,N_{t}
\equiv 0\mbox{,}\,1\mbox{,}\,\ldots \mbox{,}\, N_{t} - 1\, $; and
the momentum-space gluon propagator is studied as a function of
the magnitude of the lattice momentum
\be
p(k) \,\equiv\,\sqrt{ \,\sum_{\mu = 1}^{3} p_{\mu}^{2}(k) \,}
\,\equiv\,2\,\sqrt{ \,\sum_{\mu = 1}^{3} \sin^{2}{\left( \pi \,k_{\mu} \right)}
\,} \;\mbox{.}
\label{eq:p2}
\ee

If we define the momentum-space gluon field as
\be
{\widetilde A}_{\mu}^{b}(k)\,\equiv\,\sum_{x} \, A_{\mu}^{b}(x) \,
\exp{\left[ 2 \pi i \left(k \cdot x + k_{\mu}/2\right)\right]}
\;\mbox{,}
\ee
then eqs.\ \reff{eq:D0def} and \reff{eq:Dkdef} can be rewritten as
\ba
D(0)& \equiv& \frac{1}{9 V} \sum_{\mu\mbox{,}\,b}\,\langle\,
  \left[\,{\widetilde A}_{\mu}^{b}(0)\,\right]^{2} \rangle
\label{eq:D0def2} \\
D(k) & \equiv & \frac{1}{6 V} \sum_{\mu\mbox{,}\,b}\,\langle\,
{\widetilde A}_{\mu}^{b}(k)\,{\widetilde A}_{\mu}^{b}(-k)
\, \rangle \;\mbox{.}\;\;\;\;\;\;
\label{eq:Dkdef2}
\ea
Notice that $D(0)$ in eq.\ \reff{eq:D0def} [or in eq.\ \reff{eq:D0def2}]
is not given by $D(k)$ in eq.\ \reff{eq:Dkdef} [or in eq.\ \reff{eq:Dkdef2}]
at $k = 0$. The difference is due to
the Landau gauge condition --- the continuum-like condition as in
eq.\ \reff{eq:diverg0} --- which in momentum space reads
\be
\sum_{\mu = 1}^{3}\, p_{\mu}(k) \, {\widetilde A}_{\mu}^{b}(k)
\, =\, 0
\qquad \qquad \;\;\; \forall \; \; \; k \mbox{,} \; \; b \;\mbox{.}
\label{eq:defp}
\ee
If $k \neq (0\mbox{,}\, 0\mbox{,}\, 0)$ we obtain that only two
of the three Lorentz components of
${\widetilde A}^{b}(k)$ --- and therefore of $A^{b}(x)$ --- are independent.
This explains the factor $6$ (instead of $9$) in eqs.\ \reff{eq:Dkdef}
and \reff{eq:Dkdef2}.

Let us also note that the zero three-momentum gluon propagator $D(0)$ can be
written as
\be
D(0) \, = \, \frac{V}{9} \sum_{\mu\mbox{,}\,b}\,\langle\,
    ( {\cal A}^{b}_{\mu} )^{2} \,\rangle
\label{eq:newD0}
\;\mbox{,}
\ee
where $ {\cal A}^{b}_{\mu} \equiv \, V^{-1} \,\sum_{x}\, A^{b}_{\mu}(x) $
is the zero-momentum component of the gluon field $A_{\mu}^{b}(x)$.
Notice that a nonzero value for the constants $\,{\cal A}^{b}_{\mu}\,$
is a lattice artifact related to the use of periodic boundary conditions
and to the finiteness
of the volume. In fact, after the Landau gauge condition is imposed,
these constants are identically null --- even on a finite lattice ---
if free boundary conditions are considered
\cite{SZ},
while in the periodic case they must go to zero
in the infinite-volume limit
\cite{Z4,Z1}.
Of course, due to the volume factor in eq.\ \reff{eq:newD0}, the latter
result does not imply that $D(0)$ should be zero
in the infinite-volume limit. Nevertheless, as mentioned in the Introduction,
it has been proven by Zwanziger
\cite{Z1,Z2}
that, in this limit and in three dimensions, the
gluon propagator is less singular at momentum $p = 0$ than $p^{-1}$ and
that, very likely, it vanishes in the infrared
limit.
 
 
\section{Numerical simulations}
\label{sec:numeri}

In Table \ref{tab:runs2} we report, for each pair $(\beta\mbox{,}\,V)$, the
parameters used for the simulations.\footnote{Computations were performed on
a SUN Ultra-1 and on a SUN Ultra-2 at the Universit\`a di Roma ``Tor Vergata'',
where part of this work has been done, and on an ALPHAstation 255 at the
ZiF-Center in Bielefeld.} Overall, we have considered about 
$4100$ configurations. In all our runs we have started from a randomly
chosen lattice gauge configuration. To thermalize the gauge configuration
$\{U_{\mu}(x)\}$ we use a {\em hybrid overrelaxed} (HOR) algorithm
\cite{Teper,BW},
i.e.\ $\, m$ microcanonical (or energy-conserving)
update sweeps are done, followed
by one standard local ergodic update (heat-bath sweep) of the lattice.
In order to optimize the
efficiency of the heat-bath code, we implement two different $SU(2)$
generators (methods 1 and 2 described in Appendix A of Reference
\cite{EFGS},
with $h_{cutoff} = 2$).
In our case we did {\em not} try to find the best tuning for the value of $m$.
By analogy with the four-dimensional case
\cite{Athesis,Attilio1}
we set $m = N_{s} / 2$.

For all the pairs $(\beta\mbox{,}\, V)$,
we evaluated the {\em integrated autocorrelation time}\footnote{For a definition
see for example
\cite{S}.
To evaluate the integrated autocorrelation time we use an automatic
windowing procedure
\cite{S}
with two different window factors ($6$ and $15$). We also employ a method
\cite{zhybrid2}
based on a comparison between the naive
statistical error with a jack-knife binning error
\cite{W2}.
In all cases we checked that these three estimates are in agreement.}
$\tau_{int}$ for the Wilson loops
\be
W_{l,l} \equiv \frac{1}{3\,V}\, \frac{\Tr}{2}\,\sum_{\nu > \mu} \, \sum_{x}
\, U_{\mu\mbox{,}\nu}^{l,l}(x)
\qquad \qquad
l\,=\, 1\mbox{,}\,2\mbox{,}\,4\mbox{,}\,\ldots \mbox{,}\,N_{s} / 2
\;\mbox{,}
\label{eq:Wll}
\ee
where
\ba
U_{\mu\mbox{,}\nu}^{l,l}(x) & \equiv &
U_{\mu}(x) \,\ldots\, U_{\mu}(x + (l-1)\,e_{\mu}) \,
U_{\nu}(x + (l-1)\, e_{\mu}) \,\ldots\, U_{\nu}(x +
(l-1)\, e_{\mu} + (l-1)\,e_{\nu})
\nonumber \\
& & \,\,\, \times \,
U_{\mu}^{-1}(x + (l-1)\, e_{\mu} + (l-1)\,
e_{\nu}) \,\ldots\, U_{\mu}^{-1}(x + (l-1)\, e_{\nu}) \,
U_{\nu}^{-1}(x + (l-1)\, e_{\nu}) \,\ldots\, U_{\nu}^{-1}(x)
\;\mbox{,}
\ea
and for the Polyakov loops
\be
P_{\mu} \equiv \frac{N_{\mu}}{V}
   \sum_{\nu \neq \mu} \, \sum_{x_{\nu}} \, \frac{\Tr}{2}
     \prod_{n_{\mu} = 1}^{N_{\mu}} \,
        U_{\mu}(x + n_{\mu}\,e_{\mu})
\qquad \qquad \;
\mu\,=\, 1\mbox{,}\,2\mbox{,}\,3 \;\mbox{.}
\label{eq:Pmu}
\ee
In all cases 
we obtained $\tau_{int} \ltapprox 1$.
(Note that $\tau_{int} \,=\, 0.5$ indicates
that two successive configurations generated in the Monte Carlo simulation are
independent.) Since, for all pairs $(\beta\mbox{,}\, V)$ and for all quantities,
the number of sweeps between two consecutive configurations used for evaluating
the gluon propagator (see Table \ref{tab:runs2}) is much larger than 
the corresponding integrated autocorrelation time,
we may conclude that these configurations
are essentially statistically independent.

 
\vspace{0.3cm}
 
For the numerical gauge fixing we use the so-called {\em stochastic
overrelaxation} algorithm
\cite{FG,CM}.
In all our simulations we stop the gauge fixing when the condition
\be
\frac{1}{V} \sum_{x\mbox{,}\, b} \, \Big[
    \left( \nabla \cdot A \right)^{b}(x) \Big]^{2} \, \leq \,
     10^{- 12}
\label{eq:divergenza2}
\ee
is satisfied. [See eq.\ \reff{eq:diverg0} for the definition of the lattice
divergence $\, \left(\nabla\cdot A \right)^{b}(x) \,$ of the gluon field
$ A_{\mu}^{b}(x) $.] This is equivalent
\cite{CM}
to fixing the minimizing functional $\,{\cal E}_{U}[ g ]\,$
up to about one part in $10^{12}$. In the final gauge-fixed configuration we
also evaluate
\cite{CM}
\be
Q \, \equiv \, \frac{1}{9} \, \sum_{\nu} \frac{1}{N_{\nu}}
   \sum_{x_{\nu}\mbox{,}\, b}  \,
    \left[ \,  Q_{\nu}^{b}(x_{\nu}) - {\widehat Q}_{\nu}^{b}  \,
      \right]^{2} \, \left[ {\widehat Q}_{\nu}^{b} \right]^{- 2}
\label{eq:Qdefinizione}
\;\mbox{,}
\ee
where $ {\widehat Q}_{\nu}^{b} \, \equiv \, N_{\nu}^{-1} \sum_{x_{\nu}} \,
Q_{\nu}^{b}(x_{\nu}) $,
and $Q_{\nu}^{b}(x_{\nu})$ has been defined in eq.\ \reff{eq:charges}.
The quantity $ Q $ should be zero when the configuration is gauge-fixed, and it
is a good estimator of the quality of the gauge fixing. As in References
\cite{CM},
we found that the stochastic overrelaxation algorithm is very efficient in
fighting {\em critical slowing-down}
\cite{S},
and in making the quantities
$Q_{\nu}^{b}(x_{\nu})$ constant, i.e.\ $ Q \approx 0 $. In particular,
by averaging
over all the gauge-fixed configurations,
we find $ Q = 3.5(10)\,\times\,10^{-6}$.
We also obtain $Q \leq 10^{- 8}$ for $69 \%$ of the gauge-fixed configurations.

 
\subsection{String tension and lattice spacing}

For each coupling $\beta$ we evaluate the average plaquette
$ \langle W_{1,1} \rangle $ (see Table \ref{tab:risu}). Results for
$\beta = 5.0$ and $6.0$
are in agreement with the data reported in Table 15 of Reference
\cite{Teper}.
In Figure \ref{fig:W} we also plot $ \langle W_{1,1} \rangle $ as a function of
the coupling $\beta$, and we compare
the numerical data with the leading strong-coupling expansion $\beta / 4$
and weak-coupling expansion $\exp{(- 1 / \beta)}$. It is clear that
the crossover region from strong coupling to weak coupling occurs around
$\beta \approx 3$, in agreement with Reference
\cite{3dprimo},
and that our simulations range from the strong-coupling region up to
the weak-coupling one.

Following Reference
\cite{Teper}
we also evaluate, for $\beta \geq 3.4$,
the tadpole-improved coupling $\beta_{I} \equiv \beta\,
\langle W_{1,1} \rangle $ (see Table \ref{tab:risu}).
Then, by using the fit given in eq.\ (67) of that reference
(which is valid for $\beta \gtapprox 3.0$), we calculate the string tension
$ \sqrt{\sigma} $ in lattice units, and the inverse lattice spacing $a^{-1}$
using the input value $\sqrt{\sigma} = 0.44$ GeV (see the last two columns in
Table \ref{tab:risu}).

Finally, in Table \ref{tab:volu} we report (for each $\beta \geq 3.4$) the
lattice spacing in fm, the
largest lattice volume $V_{max}$ considered, the corresponding physical
volume in fm$^3$, and the smallest non-zero momentum (in GeV) that can be
considered for that lattice. Thus, in this work, we can explore the
infrared behavior of the gluon propagator for momenta as small as $p
\approx 110 $ MeV, in relatively large physical volumes, and for couplings
$\beta$ above the strong-coupling region. Let us notice that, if we compare
the data for the string tension (in lattice units) with data obtained for the
$SU(2)$ group in four dimensions (see for example Table 3 in Reference
\cite{Karsch}),
then our largest value of $\beta$, namely 6.0,
corresponds to $\beta \approx 2.4$ in the four dimensional case.

 
\subsection{Gribov copies}
 
In this work we do not consider
the problem of Gribov copies (see for example
\cite{Athesis,Attilio1}
and references therein.) This is motivated by our finding in the study of the
four-dimensional case. In fact, in References
\cite{Athesis,Attilio1}
we checked that, for the $SU(2)$ group in the
four-dimensional case, the influence
of Gribov copies on the gluon propagator is of the order of magnitude
of the numerical accuracy. (A similar result has also been obtained for the
Coulomb gauge
\cite{CZ}.)
In fact, from Table 2 in Reference
\cite{Attilio1},
it is clear that data corresponding to the minimal Landau gauge (absolute
minima of the functional ${\cal E}_{U}[ g ]$) are in complete agreement,
within statistical errors, with those obtained in a generic Landau gauge
(local minima of ${\cal E}_{U}[ g ]$). In particular, this seems
to be the case even at small values of the coupling $\beta$, namely
in the strong-coupling regime, where the number of Gribov copies is
higher and their effects, if present, should be larger and more easily
detectable.

 
\subsection{$Z(2)$ symmetry}
 
In References
\cite{Mitri}
it was shown for the four-dimensional case that, at {\em very large}
$\beta$, the data for the gluon propagator
are strongly affected by the broken $Z(2)$ symmetry. In particular, one
can consider all the possible combinations of signs of the average Polyakov
loops $ \langle P_{\mu} \rangle$ [see eq.\
\reff{eq:Pmu}], for a total of $2^4 = 16$ different
states. Then, if the expectation values are evaluated only over configurations
belonging to the same state, the gluon propagator takes different
values in different states
\cite{Mitri}.
Here we did the same analysis at $\beta = 5.0$
and lattice volume $V = 16^3$ with $1000$ configurations. Since we work
in three dimensions, there are $2^3 = 8$ possible states, i.e.\
combinations of the signs of the average Polyakov loops. We obtain that,
also in this case, the gluon propagator depends strongly on the state used
for evaluating the expectation value. For example, for $k = (0\mbox{,}\,
0\mbox{,}\, 1)$ the smallest value --- $D(k) = 1.50(3)$ --- is obtained
for the state characterized by positive Polyakov loops in the three directions,
while the largest value --- $D(k) = 1.68(3)$ --- corresponds to the state
characterized by negative Polyakov loops in the three directions. The two values
clearly differ by several standard deviations. A similar result is 
obtained when other momenta $k$ are considered. This observation may explain
why the data for the gluon propagator are usually characterized
by large statistical fluctuations.

 
\section{Infrared behavior of the gluon propagator}
\label{sec:infra}

In Figures \ref{fig:gluo0}--\ref{fig:gluo2} we plot the data for the
gluon propagator [see eqs.\ \reff{eq:D0def} and \reff{eq:Dkdef}] as a
function of the lattice momentum $p(k)$, defined in
eq.\ \reff{eq:p2}, for different lattice volumes $V$ and couplings $\beta$.
Our data confirm previous results
\cite{Athesis,Attilio1,Attilio2,Naka}
obtained in the strong-coupling regime for the four-dimensional
case:\footnote{~In particular see Figure 1 in Reference
\protect\cite{Attilio2}.}
the gluon propagator is decreasing as $p(k)$ decreases, provided that $p(k)$
is smaller than a value $p_{dec}$.
Also, as in four dimensions, the lattice
size at which this behavior for the gluon propagator starts to be observed
increases with the coupling $\beta$. In particular,
in the strong-coupling regime,
this propagator is clearly decreasing as $p(k)$ goes to zero,
even for relatively small lattice volumes
(see Figures \ref{fig:gluo0} and \ref{fig:gluo1}).
On the contrary, for $\beta \geq 3.4$ (see Figure \ref{fig:gluo2}),
this propagator
is increasing (monotonically) in the infrared limit for $V = 16^3$, while it
is decreasing at the largest lattice volume considered.

Let us also notice that, at high momenta, there are very small finite-size
effects, at all values of $\beta$. The situation is completely different in
the small-momenta sector, as already stressed above. Moreover, the value
$D(0)$ of the gluon propagator at zero momentum decreases monotonically as
the lattice volume increases (see for example the case $\beta = 5.0$ in Figure
\ref{fig:gluo2}). These results suggest a finite value for $D(0)$ in the
infinite-volume limit, but it is not clear whether this value would be zero or a
strictly positive constant. Therefore, the possibility of a zero value for
$D(0)$ in the infinite-volume limit is not ruled out.

Finally, in Figure \ref{fig:glukk} we plot (for
$\beta = 5.0$ and $6.0$ and $V = 64^{3}$) the data for the
gluon propagator $D(k)$ with the choice
$k = (0\mbox{,}\, 0\mbox{,}\, k_{t})$, together with the data for
$k = (1\mbox{,}\, 1\mbox{,}\, k_{t})$.
In both cases the two sets of data seem to fall on a single
curve, i.e.\ we see no sign of breaking of rotational invariance.


\vspace{0.3cm}

As said in Section \ref{sec:propag}, with our definition of the gluon
field [see eqs.\
\reff{eq:Amux} and \reff{eq:Amuxb}] the quantity $ a^{-1} A_{\mu}^{b}(x) $
approaches $( 1 / 2 ) \, g \, [ A^{(cont)} ]^{b}_{\mu}(x) \,$
in the continuum limit
$a \to 0$,  where $[ A^{(cont)} ]^{b}_{\mu}(x) $
is the (unrenormalized) conventional vector potential.
In the same limit, $\,a \,D(k)\,$ approaches $\, g^2 D^{(cont)}(k)/ 4 $,
where $\,D^{(cont)}(k)\,$ is the unrenormalized continuum gluon
propagator.
[We recall that, in three dimensions, $\,D^{(cont)}(k)\,$ has mass
dimension $-2$ and $g^2$ has mass dimension $1$.]
In Figure \ref{fig:scaling}
we plot $ \,a\,D(k)\, $ (in $\mbox{GeV}^{-1}$) as a function of the momenta
$ a^{-1} p(k) $ (in $\mbox{GeV}$) for three different values of the coupling
$\beta$: $3.4\mbox{,}\, 4.2$ and $5.0$. In all cases we
consider the largest lattice volume available (see Table \ref{tab:runs2}).
The data show good scaling in the region where finite-size effects
are negligible, i.e.\ in the limit of large
momenta (see Figure \ref{fig:scaling}).~\footnote{~We have checked
scaling, in the limit of large momenta, also for the data at
$\beta = 6.0$. However, these data are not included in Figure
\protect\ref{fig:scaling} for clarity.}
On the contrary, the scaling is poorer in the infrared limit as expected.
Nevertheless, we can see that the gluon propagator is decreasing for momenta
$p \ltapprox p_{dec}$, and that the value of $p_{dec}$ (in physical units)
is practically $\beta$-independent.
From our data (see Figure \ref{fig:scaling}b) we can set
$p_{dec} \approx 350 \, \mbox{MeV} $. Let us notice that
$p_{dec}$ corresponds to the mass scale $\gamma^{1/4}$ in a
Gribov-like propagator.

Finally, for the same set of data we consider (see Figure
\ref{fig:scaling2}) the plot of the dimensionless product $\,p(k)\,D(k)\,$
as a function of
$ a^{-1} p(k) $. Since $\,p(k)\,D(k)\,$ is decreasing for
$ a^{-1} p(k) \ltapprox 700 \, \mbox{MeV} $, we can say that the gluon
propagator is less singular than $p^{-1}$ in the infrared limit,
in agreement with Zwanziger's prediction. We notice that
the turn-over value
$p_{to} \approx 700 \, \mbox{MeV} $ is in good agreement
with the result obtained recently in four dimensions for the
$SU(3)$ group (see Figure 8 in Reference
\cite{par}).

 
\section{Conclusions}

We think that our data for the gluon propagator are very interesting.
The prediction
\cite{Z1,Gr,Z2,Stingl,smekal}
of a propagator decreasing for momenta $p \ltapprox p_{dec}$ is clearly
verified numerically for several values of the coupling $\beta$, ranging
from the strong-coupling regime to the scaling region.
Moreover, as in the four-dimensional case
\cite{Athesis,Attilio1,Attilio2},
it appears that the
lattice size at which this behavior for the gluon propagator starts to be
observed increases with the coupling.
This requirement of large lattice volumes could explain why
a decreasing gluon propagator
has never been observed in the scaling region for the four-dimensional
case
\cite{MO2,Gup,Nakamura,Aiso,gut,par,Athesis,Attilio1,Attilio2,Naka}.

As mentioned above, our data in the strong-coupling regime and in
three dimensions are in qualitative
agreement with results obtained at small $\beta$ in the four dimensional
case
\cite{Athesis,Attilio1,Attilio2,Naka}.
This strongly suggests to us that a similar analogy will hold
--- in the limit of large lattice volumes --- for couplings $\beta$
in the scaling region, leading to an infrared-suppressed gluon
propagator also in the four-dimensional case.

 
\section*{Acknowledgments}
I would like to thank T.\ Mendes and D.\ Zwanziger for valuable discussions
and suggestions. I also thank R.\ Alkofer and A.\ Hauck for e-mail
correspondence. This work was partially supported by the TMR network
Finite Temperature Phase Transitions in Particle Physics,
EU contract no.: ERBFMRX-CT97-0122.


\clearpage

\begin{figure}[p]
\begin{center}
\epsfxsize=0.40\textwidth
\leavevmode\epsffile{Plot_w11_axis.ps}
\end{center}
\vspace*{2.5cm}
\caption{~Plot of the average plaquette
$\langle W_{1,1} \rangle$ as a function of the coupling $\beta$.
For comparison we also plot the leading strong-coupling expansion $\beta / 4$
and weak-coupling expansion $\exp{(- 1 / \beta)}$. Error bars are not
visible.
}
\label{fig:W}
\end{figure}

\begin{figure}[p]
\begin{center}
\epsfxsize=0.40\textwidth
\leavevmode\epsffile{Plot_Vgluon00_axis_bis.ps}
\end{center}
\vspace*{2.5cm}
\caption{~Plot of the lattice gluon propagator $D(k)$
[see eqs.\ \protect\reff{eq:D0def}
and \protect\reff{eq:Dkdef}] as a function of the lattice momentum
$p(k)$ [see eq.\ \protect\reff{eq:p2}] for lattice volumes $V = 16^{3}$
($\Box$) and $V = 32^{3}$ ($\ast$), with $k =
(0\mbox{,}\, 0\mbox{,}\, k_{t})$, at $\beta = 0$.
Error bars are one standard deviation.
}
\label{fig:gluo0}
\end{figure}

\clearpage

\begin{figure}[p]
\begin{center}
\vspace*{0cm}
\epsfxsize=0.40\textwidth
\protect\hspace*{0.1cm}
\leavevmode\epsffile{Plot_Vgluon10_axis_bis.ps}
\hspace*{1.4cm}
\epsfxsize=0.40\textwidth
\epsffile{Plot_Vgluon16_axis_bis.ps}  \\
\vspace*{1.5cm}
\epsfxsize=0.40\textwidth
\protect\hspace*{0.1cm}
\leavevmode\epsffile{Plot_Vgluon22_axis_bis.ps}
\hspace*{1.4cm}
\epsfxsize=0.40\textwidth
\epsffile{Plot_Vgluon28_axis_bis.ps}  \\
\end{center}
\vspace*{2.5cm}
\caption{~Plot of the lattice gluon propagator $D(k)$
[see eqs.\ \protect\reff{eq:D0def}
and \protect\reff{eq:Dkdef}] as a function of the lattice momentum
$p(k)$ [see eq.\ \protect\reff{eq:p2}] for lattice volumes $V = 16^{3}$
($\Box$) and $V = 32^{3}$ ($\ast$), with $k =
(0\mbox{,}\, 0\mbox{,}\, k_{t})$, at:
     ({\bf a}) $\beta = 1.0$;
     ({\bf b}) $\beta = 1.6$;
     ({\bf c}) $\beta = 2.2$;
     ({\bf d}) $\beta = 2.8$.
Error bars are one standard deviation.
}
\label{fig:gluo1}
\end{figure}

 
\clearpage

\begin{figure}[p]
\begin{center}
\vspace*{0cm}
\epsfxsize=0.40\textwidth
\protect\hspace*{0.1cm}
\leavevmode\epsffile{Plot_Vgluon34_axis_bis.ps}
\hspace*{1.4cm}
\epsfxsize=0.40\textwidth
\epsffile{Plot_Vgluon42_axis_bis.ps}  \\
\vspace*{1.5cm}
\epsfxsize=0.40\textwidth
\protect\hspace*{0.1cm}
\leavevmode\epsffile{Plot_Vgluon50_axis_bis.ps}
\hspace*{1.4cm}
\epsfxsize=0.40\textwidth
\epsffile{Plot_Vgluon60_axis_bis.ps}  \\
\end{center}
\vspace*{2.5cm}
\caption{~Plot of the lattice gluon propagator $D(k)$
[see eqs.\ \protect\reff{eq:D0def}
and \protect\reff{eq:Dkdef}] as a function of the lattice momentum
$p(k)$ [see eq.\ \protect\reff{eq:p2}] for lattice volumes $V = 16^{3}$
($\Box$), $V = 16^{2}$x$32$ ($+$), $V = 32^{3}$ ($\ast$), $V = 32^{2}$x$64$
($\circ$), and $V = 64^{3}$ ($\Diamond$), with $k = (0\mbox{,}\, 0\mbox{,}\,
k_{t})$, at:
     ({\bf a}) $\beta = 3.4$;
     ({\bf b}) $\beta = 4.2$;
     ({\bf c}) $\beta = 5.0$;
     ({\bf d}) $\beta = 6.0$.
Error bars are one standard deviation.
}
\label{fig:gluo2}
\end{figure}
 
 
\clearpage

\begin{figure}[p]
\begin{center}
\vspace*{0cm}
\epsfxsize=0.40\textwidth
\protect\hspace*{0.1cm}
\leavevmode\epsffile{Plot_64gluon50_axis_large_new2_bis.ps}
\hspace*{1.4cm}
\epsfxsize=0.40\textwidth
\epsffile{Plot_64gluon60_axis_large_new2_bis.ps}  \\
\end{center}
\vspace*{2.5cm}
\caption{~Plot of the lattice gluon propagator $D(k)$
[see eqs.\ \protect\reff{eq:D0def}
and \protect\reff{eq:Dkdef}] as a function of the
lattice momentum $p(k)$ [see eq.\ \protect\reff{eq:p2}] for lattice volume
$V = 64^{3}$, with $k = (0\mbox{,}\, 0\mbox{,}\, k_{t})$ ($\ast$), and
$k = (1\mbox{,}\, 1\mbox{,}\, k_{t})$ ($\Box$), at:
     ({\bf a}) $\beta = 5.0$;
     ({\bf b}) $\beta = 6.0$.
Error bars are one standard deviation.
}
\label{fig:glukk}
\end{figure}


\begin{figure}[p]
\begin{center}
\epsfxsize=0.40\textwidth
\leavevmode\epsffile{Plot_gluon_resc_34_42_50_axis_bis.ps}
\hspace*{1.4cm}
\epsfxsize=0.40\textwidth
\epsffile{Plot_gluon_rescbis_34_42_50_axis_bis.ps}  \\
\end{center}
\vspace*{2.5cm}
\caption{~Plot of the gluon propagator $a \, D(k)$ in physical units
($\mbox{GeV}^{-1}$) as a function of the
momenta $a^{-1} \, p(k)$ (in $\mbox{GeV}$)
with $k = (0\mbox{,}\, 0\mbox{,}\, k_{t})$ and
$k = (1\mbox{,}\, 1\mbox{,}\, k_{t})$, for
$\beta = 3.4$ and $V = 32^3$ ($\Box$), $\beta = 4.2$ and $V = 64^3$ ($\ast$),
$\beta = 5.0$ and $V = 64^3$ ($\Diamond$).
In the second figure only the infrared
region $a^{-1} \, p(k) \protect\ltapprox 1 \mbox{GeV}$ is considered.
Error bars are one standard deviation.}
\label{fig:scaling}
\end{figure}

\clearpage

 
\begin{figure}[p]
\begin{center}
\epsfxsize=0.40\textwidth
\leavevmode\epsffile{Plot_gluon_pD_34_42_50_axis_bis.ps}
\hspace*{1.4cm}
\epsfxsize=0.40\textwidth
\epsffile{Plot_gluon2_pD_34_42_50_axis_bis.ps} \\
\end{center}
\vspace*{2.5cm}
\caption{~Plot of the dimensionless product
$p(k) \, D(k)$ as a function of the
momenta $a^{-1} \, p(k)$ (in $\mbox{GeV}$)
with $k = (0\mbox{,}\, 0\mbox{,}\, k_{t})$ and
$k = (1\mbox{,}\, 1\mbox{,}\, k_{t})$, for
$\beta = 3.4$ and $V = 32^3$ ($\Box$), $\beta = 4.2$ and $V = 64^3$ ($\ast$),
$\beta = 5.0$ and $V = 64^3$ ($\Diamond$).
In the second figure only the infrared
region $a^{-1} \, p(k) \protect\ltapprox 1 \mbox{GeV}$ is considered.
Error bars are one standard deviation.}
\label{fig:scaling2}
\end{figure}


\clearpage

\widetext
\begin{table}
\caption{
The pairs $(\beta\mbox{,} V)$ used for the simulations, the number of
configurations, the number of HOR sweeps used for thermalization,
the number of HOR sweeps between two consecutive
configurations used for evaluating the gluon propagator, and the parameter
$p_{so}$
used by the stochastic overrelaxation algorithm.}
\label{tab:runs2}
\begin{tabular}{ c c c c c c }
 $\beta$ & $V$ & Configurations & Thermalization & Sweeps & $p_{so}$ \\ \hline
0.0 & $16^3$ &  200 &    4 &   2 & 0.85 \\
0.0 & $32^3$ &  100 &    4 &   2 & 0.90 \\ \hline
1.0 & $16^3$ &  200 & 1100 & 100 & 0.84 \\
1.0 & $32^3$ &  100 & 1100 & 100 & 0.93 \\ \hline
1.6 & $16^3$ &  200 & 1650 & 150 & 0.81 \\
1.6 & $32^3$ &  100 & 1650 & 150 & 0.83 \\ \hline
2.2 & $16^3$ &  200 & 2200 & 200 & 0.75 \\
2.2 & $32^3$ &  100 & 2200 & 200 & 0.76 \\ \hline
2.8 & $16^3$ &  200 & 2750 & 250 & 0.72 \\
2.8 & $32^3$ &  100 & 2750 & 250 & 0.75 \\ \hline
3.4 & $16^3$ &  200 & 3025 & 275 & 0.69 \\
3.4 & $32^3$ &  100 & 3025 & 275 & 0.72 \\ \hline
4.2 & $16^3$ &  200 & 3300 & 300 & 0.66 \\
4.2 & $32^3$ &  100 & 3300 & 300 & 0.70 \\
4.2 & $64^3$ &   56 & 3300 & 300 & 0.72 \\ \hline
5.0 & $16^3$ & 1000 & 3575 & 325 & 0.63 \\
5.0 & $16^2$x$32$ &  200 & 3575 & 325 & 0.65 \\
5.0 & $32^3$ &  170 & 3325 & 325 & 0.68 \\
5.0 & $32^2$x$64$ &  100 & 3575 & 325 & 0.65 \\
5.0 & $64^3$ &   54 & 2275 & 325 & 0.69 \\ \hline
6.0 & $16^3$ &  200 & 3850 & 350 & 0.61 \\
6.0 & $32^3$ &  150 & 3350 & 350 & 0.67 \\
6.0 & $64^3$ &   97 & 3850 & 350 & 0.71
\end{tabular}
\end{table}

\widetext
\begin{table}
\caption{
For each coupling $\beta$ we report the value of the average plaquette
$ \langle W_{1,1} \rangle $, together with the volume $V$ and the number of
HOR sweeps used for the analysis.
Also, for $\beta \geq 3.4$, we report the tadpole-improved coupling $\beta_{I}$,
the string tension $ \protect\sqrt{\sigma} $ in lattice
units, and the inverse lattice spacing $a^{-1}$ in GeV. Error bars for
the string tension and the inverse lattice spacing come from propagation
of errors. Error bars for $ \langle W_{1,1} \rangle $ are one standard
deviation, evaluated taking into account the value of the integrated
autocorrelation time $\tau_{int,W_{1,1}}$ for the plaquette, namely the
variance is multiplied by $2 \tau_{int,W_{1,1}}$ (see eq.\ (3.7) in Reference 
\protect\cite{S}).
}
\label{tab:risu}
\begin{tabular}{ c c c c c c c }
 $\beta$ & $V$ & Sweeps & $\langle W_{1,1} \rangle$ & $\beta_{I}$ &
  $\sqrt{\sigma}$ & $a^{-1}\,\mbox{(GeV)}$ \\ \hline
0.0 & $32^3$ &   198 & 0.000110(113)&           &
                                                           & \\ 
1.0 & $32^3$ &  9900 & 0.241650(16) &           &
                                                           & \\ 
1.6 & $32^3$ & 14850 & 0.373147(13) &           &
                                                           & \\ 
2.2 & $32^3$ & 19000 & 0.493302(12) &           &
                                                           & \\ 
2.8 & $32^3$ & 24500 & 0.595483(11) &           &
                                                           & \\ 
3.4 & $32^3$ & 26950 & 0.672720(9) & 2.28725(3) &
                                         0.506(13) & 0.87(2) \\ 
4.2 & $64^3$ & 14149 &  0.741862(4) & 3.11582(2) &
                                          0.387(8) & 1.14(2) \\ 
5.0 & $64^3$ & 16164 &  0.786877(3) & 3.93438(1) &
                                          0.314(5) & 1.40(2) \\ 
6.0 & $64^3$ & 25135 &  0.824783(3) & 4.94870(2) &
                                            0.254(4) & 1.73(2)  
\end{tabular}
\end{table}

\widetext
\begin{table}
\caption{
For each coupling $\beta \geq 3.4$ we report the
lattice spacing in $\mbox{fm}$, the largest lattice
volume $V_{max}$, the corresponding physical volume
in $\mbox{fm}^3$,
and the smallest non-zero momentum that can be considered for 
that lattice (in GeV).
}
\label{tab:volu}
\begin{tabular}{ c c c c c }
 $\beta$ & $a\,\mbox{(fm)}$ & $V_{max}$ & $a^{3}\,V_{max}\,\mbox{(fm}^3)$ &
   $a^{-1} \,p_{min}\,\mbox{(GeV)}$ \\ \hline
3.4 & 0.226(6) & $32^3$ & $7.2^3$ 
                                & 0.171 \\
4.2 & 0.173(3) & $64^3$ & $11.1^3$ 
                                & 0.112 \\
5.0 & 0.140(2) & $64^3$ & $9.0^3$ 
                                & 0.137 \\
6.0 & 0.114(1) & $64^3$ & $7.30^3$ 
                                & 0.170
\end{tabular}
\end{table}

\end{document}